\documentclass[twocolumn,tighten,times]{aastex62}

\usepackage{amssymb,amsmath}
\usepackage[inline]{enumitem} 
\usepackage{soul}
\usepackage{float}

\newcommand{\HII}{H {\small{II}} }

\shorttitle{}
\shortauthors{Liang et al.}

\begin{document}

\title{Investigating the physical properties and fragmentation of the AFGL 333-Ridge}

\correspondingauthor{Jin-Long Xu}
\email{xujl@bao.ac.cn}

\author{Xiaolian Liang}
\affil{National Astronomical Observatories, Chinese Academy of Sciences, Beijing 100101, People's Republic of China}
\affil{University of Chinese Academy of Sciences, Beijing 100049, People's Republic of China}

\author{Jin-Long Xu}
\affil{National Astronomical Observatories, Chinese Academy of Sciences, Beijing 100101, People's Republic of China}

\author{Jun-Jie Wang}
\affil{National Astronomical Observatories, Chinese Academy of Sciences, Beijing 100101, People's Republic of China}

\begin{abstract}
We present multi-wavelength data to investigate the physical properties and fragmentation of AFGL 333-Ridge. A statistical analysis of velocity dispersion indicates that turbulence is the dominant motion in the ridge. However, the linear mass density (1124.0 $M_{\odot}/{\rm {pc}}$) of AFGL 333-Ridge far exceeds its critical value of 406.5 $M_{\odot}/{\rm {pc}}$, suggesting that additional motions are required to prevent the filament radial collapse. Using the {\it {\it getsources}} algorithm, we identified 14 cores from the {\it {\it Herschel}} maps, including two protostellar cores and 12 starless cores. All of these starless cores are gravitationally bound, and are therefore considered to be prestellar cores. Based on their radius-mass relation, 11 of 14  cores have the potential to form massive stars. Moreover, the seven cores in two sub-filaments of AFGL 333-Ridge seem to constitute two necklace-like chains with a spacing length of 0.51 pc and 0.45 pc, respectively. Compared the spacing length with theoretical prediction lengths by Jeans and cylindrical fragmentations, we argued that the combination of turbulence and thermal pressure may lead to the fragmentation of the two sub-filaments into the cores.
\end{abstract}


\section{Introduction}
It is generally accepted that stars form primarily in molecular clouds when the over-density structures within them become unstable and gravitational collapse. However, the early phase of star formation is still not well understood. The Herschel Gould Belt survey has found that most prestellar cores, which are typically considered to be the early stage of stars, are embedded in filaments with a column density $N_{\rm H_2} > 7 \times 10^{21}$ cm$^{-2}$ \citep{2010A&A...518L.102A,2015A&A...584A..91K,2020A&A...635A..34K}, implying that the filament may play an essential role in core formation. Investigating how filaments fragment into cores during the early evolutionary stages and studying the cores' properties can help gain insights into the link between filament structure and the formation of cores.

AFGL 333-Ridge is one of three active star-forming regions in W3 Giant Molecular Cloud (GMC). Moreover, AFGL 333-Ridge shows a filamentary structure and is associated with three \HII regions \citep{2021ApJ...913...14L}, three IRAS sources, and two H$_2$O masers \citep{2017PASJ...69...16N}. \cite{2021ApJ...913...14L} proposed that AFGL 333-Ridge is formed by the collision and combination of two molecular clouds with different velocity components. The majority of young stellar objects (YSOs) are found to be distributed within the collision region. The vigorous star-forming activity and high column density in AFGL 333-Ridge make the ridge an ideal target for understanding how filament fragments into cores.

In this paper, we will focus on the physical properties and fragmentation of AFGL 333-Ridge. In Section 2, we describe the multi-wavelength observations and data processing. In Section 3, we present the data analysis, including the identification of dense cores and alculation of their physical properties. In Section 4, we discuss the properties of dense cores and fragmentation of filaments. Finally, in Section 5, we provide a summary and conclusions.
 
\begin{figure*}
    \centering
    \includegraphics[width=0.99\textwidth]{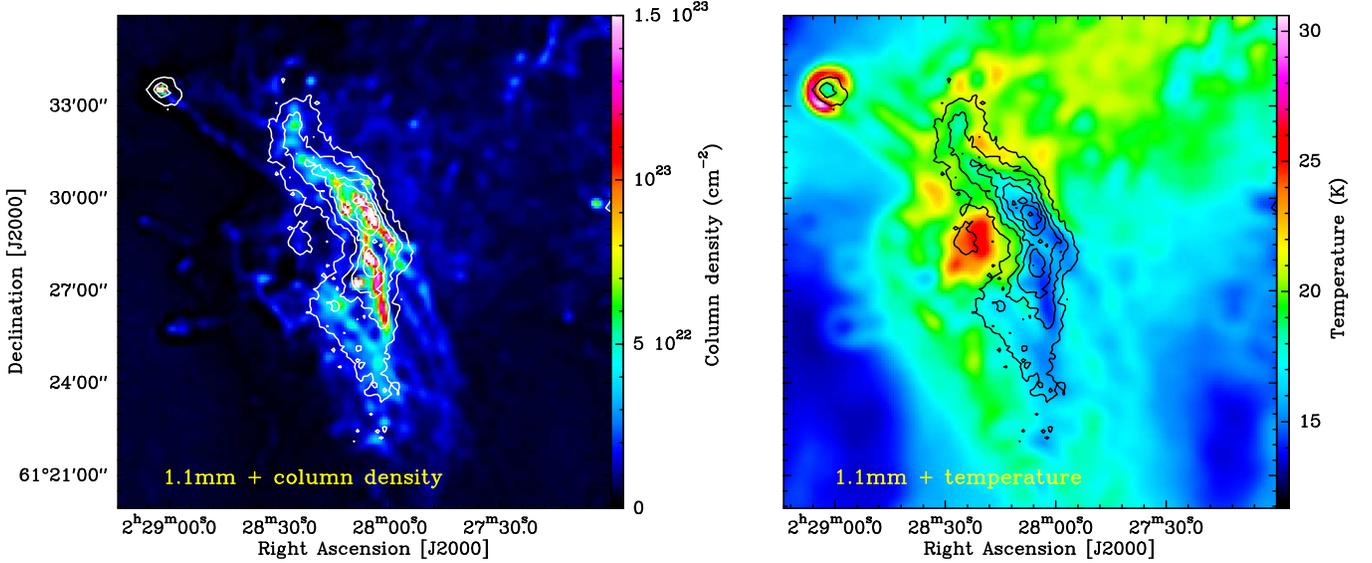}
    \caption{Left panel: the Herschel column density map superposed by the BGPS 1.1 mm continuum emission in white contours. Right panel: the Herschel dust temperature map overlaid by the BGPS 1.1 mm continuum emission in black contours. The above two maps are adapted from \cite{2021ApJ...913...14L}. }
    \label{fig:col_density}
\end{figure*}
   
\section{Observation and data reduction}
The {\it {\it Herschel}} imaging observations of Hi-GAL survey, include PACS 70 $\mu$m and 160 $\mu$m \citep{2010A&A...518L...2P} and SPIRE 250 $\mu$m, 350 $\mu$m, and 500 $\mu$m \citep{2010A&A...518L...3G}. The beam sizes of the PACS data at 70 $\mu$m and 160 $\mu$m are 8.4$^{\prime\prime}$ and 13.5$^{\prime\prime}$, respectively. The beam sizes of the SPIRE data at 250, 350 and 500 µm are 18.2$^{\prime\prime}$, 24.9$^{\prime\prime}$ and 36.3$^{\prime\prime}$, respectively. Using the method devised in Appendix A of \cite{2013A&A...550A..38P}, we created a high-resolution H$_2$ column density map by pixel-to-pixel SED fitting to {\it {Herschel}} 160 to 500 $\mu$m data with modified blackbody function.

The NARO VLA Sky Survey (NVSS) 21 cm (1.4 GHz) continuum emission data was used to trace the ionizing gas \citep{1998AJ....115.1693C}. We also used the 1.1 mm continuum data of Bolocam Galactic Plane Survey (BGPS) \citep{2011ApJS..192....4A} to trace the dense part of our target source. The C$^{18}$O $J$=1-0 (109.782 GHz) molecular line data was obtained using the PMO 13.7m radio telescope at De Ling Ha in the western of China. \cite{2021ApJ...913...14L} gave more details of the PMO 13.7 m radio telescope instrumentation. We also use the GILDAS/CLASS \footnote{http://www.iram.fr/IRAMFR/GILDAS/} package to reduce the final data.

\section{Results}
\label{sec:results}

\subsection{The physical properties of AFGL 333-Ridge}
\label{subsec:phy}
Figure~\ref{fig:col_density} shows the H$_2$ column density map and dust temperature map of AFGL 333-Ridge, which are constructed by \cite{2021ApJ...913...14L}. These maps were created by pixel-to-pixel SED fitting over a wavelength range of {\it {\it Herschel}} 160 to 500 $\mu$m with a modified blackbody function. The 1.1 mm continuum emission (black or white contours show in Figure~\ref{fig:col_density}) represents the dense region of AFGL 333-Ridge. This region is referred to as the ``ridge", which has an average column density of 4.0 $\times$10 $^{22}$ cm$^{-2}$ and an average dust temperature of 16 K \citep{2021ApJ...913...14L}. Furthermore, the ridge seems to contain sub-filaments.  However, the central part has a column density of up to 3.0 $\times$10 $^{23}$ cm$^{-2}$ and therefore has a great potential for massive star formation based on the massive star formation threshold \citep{2008Natur.451.1082K}. By the average column density ($N_{\rm H_2}$), the mass of the ridge can be estimated from  $M_{\rm H_{2}} = S\mu_gm(\rm H_{2})\it N_{\rm H_{2}}$, where $\mu_g$ is the mean atomic weight of the gas, $m(\rm H_{2})$ is the mass of a hydrogen molecule, and $S$ is the area of ridge which is measured from the 1.1mm emission contours in Figure~\ref{fig:col_density}. Finally, the mass of the ridge was calculated to be about 6.7$\times 10^3$ M$_{\odot}$.

\begin{figure}
    \centering
    \includegraphics[width=0.46\textwidth]{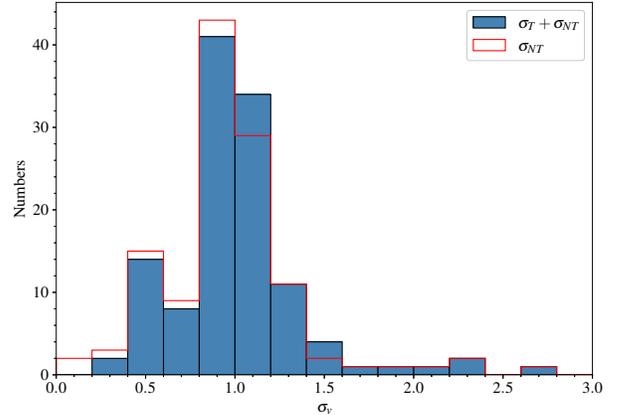}
    \vspace{-4mm}
    \caption{
      Distributions of  thermal velocity dispersion and nonthermal velocity dispersion in AFGL 333-Ridge.
    }
    \label{fig:dis}
\end{figure}

\begin{figure*}
    \centering
    \includegraphics[width=0.9\textwidth]{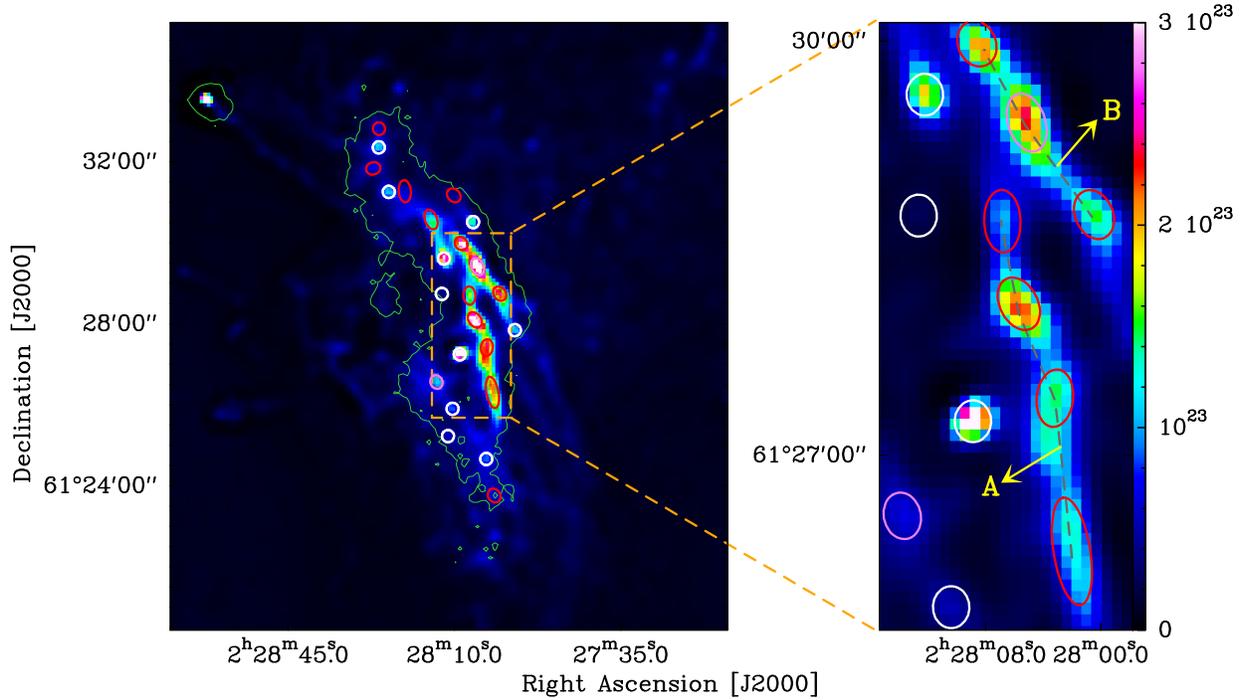}
    \caption{ Left panel: Position of the 24 dense cores overlaid on the {\it\it{Herschel}} high-resolution column density map. The violet and red ellipses represent protostellar cores, and starless cores, respectively. The 10 white ellipses mark those dense cores that cannot be accurately distinguished. The green contour is BGPS 1.1mm continuum emission is shown with the level of 0.2 Jy/beam. The orange dashed box represents the cores in the sub-filaments. Right panel: The zoomed image of candidate cores in the orange rectangle. The grey dashed lines linked the central position of seven dense cores.
    }
    \label{fig:position}
\end{figure*}  

Velocity dispersion can reflect the dynamical state of molecular clouds. The C$^{18}$O molecular line is considered to be a good tracer of dense gas, thus can be used to estimate the velocity dispersion. The 1D velocity dispersion is composed of thermal and non-thermal velocity dispersion, $\sigma_{\nu} = \sqrt{\sigma_{\rm Thermal}^2 + \sigma_{\rm NT}^2}$. The non-thermal velocity dispersion is determined by $ \sigma_{\rm NT} = (\sigma_{\rm {C^{18}O}}^2-\frac{kT_{\rm {kin}}}{m_{\rm {C^{18}O}} \mu })^{1/2}$, where $\sigma_{\rm {C^{18}O}}$ is the observed velocity dispersion of C$^{18}$O J=1-0. The $\sigma_{\rm {C^{18}O}}$ is related to the observed spectral linewidth ($\Delta V_{18}$), which can be calculated by $\sigma_{\rm {C^{18}O}} = (\Delta V_{18}/\sqrt{8{\rm {ln2}}})$. Meanwhile, thermal velocity dispersion can be estimated by $\sigma_{\rm Thermal} = \sqrt {\frac{kT_{\rm {kin}}}{m_{\rm H_2} \mu}}$, where $T_{\rm {kin}}$ is the kinetic temperature. The $T_{\rm {kin}}$ can be replaced by $T_{\rm {ex}}$, when the density of the molecular clouds is high enough and the molecular clouds is under the LTE condition \citep{2017A&A...606A.102F}. In the ridge region, we use $T_{\rm {ex}}$ instead of $T_{\rm {kin}}$, which is derived from the $^{12}$CO J=1-0 data of \cite{2021ApJ...913...14L}, ranging from 4 K to 35 K with a mean value of 22 K. Here we choose the C$^{18}$O J=1-0 emission above 3$\sigma$ to derive these physical parameters. The mean thermal 1D velocity dispersion is 0.27 km s$^{-1}$, while the nonthermal 1D velocity dispersion averaged 0.98 km s$^{-1}$.  Finally, the mean 1D velocity dispersion at the ridge is calculated to be 1.0 km s$^{-1}$. From the above calculations, we plotted the distribution of the velocity dispersion in Figure~\ref{fig:dis}.

\begin{table*}
\begin{center}
\tabcolsep 1.2mm\caption{Physical parameters of the cores identified with {\it{\it Herschel}} in the AFGL 333-Ridge}
\def\temptablewidth{1\textwidth}
\begin{tabular}{lcccccccccccccc}
\hline\hline
Number  & RA & Dec & $\rm {FWHMX}$ &$\rm {FWHMY}$&PA&$R_{\rm {eff}}$&$N_{\rm {H_2}}$&$T_{\rm {dust}}$&$M_{\rm {core}}$&$M_{\rm {BE}}$&Type  \\
  &(J2000) & (J2000) & ($^{\prime \prime}$) & ($^{\prime \prime}$)&($^{\circ}$)&(pc)&(10$^{22}$ cm$^{-2}$)&(k)&($M_{\odot}$)&($M_{\odot}$)& \\   
    \hline\noalign{\smallskip} 
\ 01  & 02:28:05.7&61:28:05.6& 25.2&18.2&38.0&0.11&26.2&12.7&225.9&13.7&prestellar core\\
\ 02 & 02:28:04.5& 61:29:58.7&21.1&18.2&43.5&0.07&23.6&14.2&84.3&9.8&prestellar core \\
\ 03 & 02:28:00.3& 61:29:24.4&26.5&18.2&25.4&0.12&25.1&12.6&256.0&14.7& prestellar core\\
\ 04 & 02:28:01.5& 61:27:24.6& 25.1&18.2& 172.4&0.11&15.3&13.6&129.7&14.5&protostellar core\\
\ 05 & 02:28:07.1&61:30:34.3&30.9&18.2&21.2&0.15&9.6&15.0&150.5&21.8&prestellar core\\
\ 06 & 02:28:03.3&61:28:44.4&22.9&18.2&37.3&0.09&15.7&13.3&91.1&11.7&prestellar core\\
\ 07 & 02:28:00.2&61:26:18.0&47.5&18.2&10.8&0.23&11.3&12.7&408.0&28.1&prestellar core\\
\ 08 & 02:28:00.3&61:28:41.6&27.2&18.2&2.8&0.13&8.2&15.9&90.6&19.4&prestellar core\\
\ 09 & 02:28:06.0& 61:26:33.5&20.7&18.2&26.5&0.06&8.4&14.5&25.7&9.4&protostellar core\\
\ 10 & 02:28:04.0&61:31:16.2&33.3&18.2&4.3&0.16&5.5&16.0&101.4&25.4&prestellar core\\
\ 11 & 02:28:25.9&61:32:48.7&19.0&18.2&17.6&0.04&5.5&14.8&5.4&5.4&prestellar core\\
\ 12 & 02:28:27.1&61:31:50.0&22.0&18.2&106.3&0.08&5.9&15.2&27.6&12.1&prestellar core\\
\ 13 & 02:28:04.6&61:23:45.3&21.8&18.2&36.4&0.08&4.4&14.6&19.4&11.2&prestellar core\\
\ 14 & 02:28:01.4&61:31:10.0&22.9&18.2&46.5&0.09&3.1&17.9&18.0&15.8&prestellar core\\
\noalign{\smallskip}\hline
\end{tabular}
\end{center}
Note:  RA and DEC are the equatorial centroid coordinates of the identified cores. FWHMX and FWHMY are the major and minor axes of each core in the {\it {\it Herschel}} high-resolution (18.2$^{\prime \prime}$) density map. PA is the position angle of the major axis. $R_{\rm {dec}}$ is the deconvolved core radius. $N_{\rm {H_2}}$ and $T_{\rm {dust}}$ are the mean column density and temperature of each core estimated from the column density map and temperature map within the derived radius. $M_{\rm {core}}$ is the core mass. $M_{\rm {BE}}$ is the critical Bonnor-Ebert (BE) mass. Type represents different kinds of dense cores. Location illustrates the different positions of the cores.
\label{table:cores}
\end{table*}

\subsection{Sources extraction from Herschel map}
The dense cores in AFGL 333-Ridge were extracted using {\it {\it getsources}}, a sources identification algorithm. This algorithm was designed mainly to identify compact objects in far-infrared or submillimeter surveys of Galactic molecular clouds. The extraction method of {\it {\it getsources}} can be divided into “detection” and “measurement” stages \citep{2012A&A...542A..81M}. To extract starless cores, we use 160 to 500 $\mu$m {\it {\it Herschel}} maps and a high-resolution (18.2$^{\prime\prime}$) column density map as an additional wavelength in the detection stage. Adding the high-resolution column density map can ensure that  the identified cores are associated with the column density peaks. For the extraction of protostellar cores or YSOs, we only used the 70 $\mu$m Herschel map in the detection stage. Because the dust around YSOs and protostellar cores will be heated to a higher temperature than the cold cores, which then makes them visible as point-like objects at 70 $\mu$m \citep{2008ApJS..179..249D}.  For both sets of extractions at the measurement stage, all five {\it {Heschel}} maps and a high-resolution column density map were used to detect the cores' position and measure the flux.

Following the criteria given by \cite{2015A&A...584A..91K}, a total of 24 cores were selected. Among them, 10 cores have the major and minor axis of 18.2$^{\prime\prime}$, which is equal to the resolution of the high-resolution column density map. Therefore, in this paper, we will only focus on the other 14 dense cores due to the beam effect. These 14 dense cores consist of 2 protostellar cores and 12 starless cores. Figure~\ref{fig:position} shows the spatial distribution of all the cores overlapped on the column density map. The white ellipses in Figure~\ref{fig:position} mark those dense cores that cannot be accurately distinguished, the red and violet ellipses represent starless cores, and protostellar cores, respectively.  We find that all the detected dense cores are distributed along the main ridge. Especially, the distributions of seven cores in the two sub-filaments seem to form two necklace-like chains, which are indicated by  two gray dashed lines in the right panel of Figure~\ref{fig:position}. We denote these two sub-filaments as sub-filament A and sub-filament B. Based on the physical information generated by {\it {\it getsources}}, as well as the column density and temperature maps, we could calculate the relevant physical parameters for each core. The effective core radius is defined as the mean deconvolved FWHM radius of equivalent elliptical Gaussian sources, which can be calculated as $R_{\rm {eff}}=\sqrt{FWHMX \times FWHMY-HPBW^2} $ \citep{2015A&A...584A..91K}. The $FWHMX$ and $FWHMY$ denoted that the angular size of the major axis and minor axis of the cores, and $HPBW$  is the angular resolution (18.2$^{\prime \prime}$) of the column density map. Similarly, The masses of the cores can be derived by $M_{\rm {core}}$ = $\pi R^2 \mu_g m(H_2)N_{\rm {H_2}}$. The $N_{\rm {H_2}}$ is the average column density of the cores, which can be estimated from the integrated column density within the deconvolved radius from the column density map. All the estimated physical parameters of the 14 dense cores are listed in Table~\ref{table:cores}.

\section{Discussion}
\label{dis}
\subsection{Stability of AFGL 333-Ridge}
In Section~\ref{subsec:phy}, we have calculated the velocity dispersion for AFGL 333-Ridge. On the basis of this calculation, we made statistics on the distribution of velocity dispersion, show in Figure~\ref{fig:dis}. We found that the nonthermal motion dominates the velocity dispersion. The nonthermal motion is generally interpreted as supersonic turbulence \citep{2007ARA&A..45..565M}. The critical linear mass is a threshold, above which the filament becomes gravitationally unstable, contracted radially, and fragmented along its length \citep{1992ApJ...388..392I,1997ApJ...480..681I}. When turbulent pressure overtakes thermal pressure in the filament, the critical linear mass can be written as $(M/l)_{line,crit} = 84(\Delta V)^2 M_{\odot} pc^{-1}$, where $\Delta V$ is in units of km s$^{-1}$ \citep{2010ApJ...719L.185J}. The typical linewidth observed in the AFGL 333-Ridge is 2.2 km s$^{-1}$, then calculated $(M/l)_{line,crit}$ as 406.5 $M_{\odot}/{\rm {pc}}$. Depending on the mass (6.7$\times 10^3$ M$_{\odot}$) and length (6.0 pc) of the ridge, we estimate that the observed linear mass of the ridge is 1124.0 $M_{\odot}/{\rm {pc}}$, which is much greater than the critical linear mass. Therefore, we assume that in addition to the turbulence, AFGL 333-Ridge also needs other motions, such as magnetic field or feedback, to help stabilize it against radial collapse. 

\subsection{Properties and formation of cores in AFGL 333-Ridge}
In the AFGL 333-Ridge, we have identified 14 candidate cores, comprising 2 protostellar cores and 12 starless cores. The starless cores which are self-gravitating are likely to form stars in the future. Using the critical Bonnor-Ebert mass, we can determine whether starless cores are self-gravitating or not. When the $\alpha_{\rm BE}$ = $M_{\rm {BE,crit}}/M_{\rm core} \leq 2$, the starless core can be considered as a self-gravitating prestellar core \citep{2007prpl.conf...33W}. The critical Bonnor-Ebert (BE) mass can be expressed as $M_{\rm {BE,crit}}\approx$ 2.4$R_{\rm BE} c_s^2/G$, where $R_{\rm {BE}}$ is the BE radius, $G$ is the gravitational constant, and $c_{\rm s}$ is the sound speed. Estimating the BE mass ($M_{\rm {BE}}$) of each object by using the deconvolved core radius ($R_{\rm dec}$) to replace the BE radius, and the outcomes are listed in Table~\ref{table:cores}. After calculation and comparison, all 12 starless cores are considered to be bound prestellar cores.

 \begin{figure}
    \centering
    \includegraphics[width=0.47\textwidth]{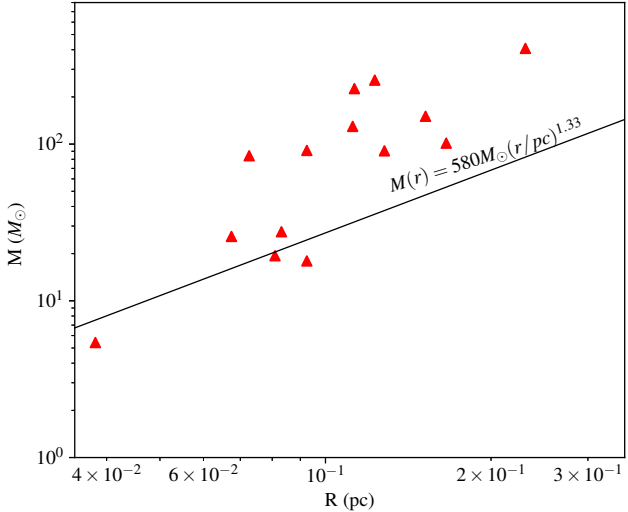}
    \caption{Mass versus radius diagram for all the candidate cores in AFGL 333-Ridge. The solid black line represents the threshold \citep{2010ApJ...723L...7K} that can separate whether massive stars can form or not.}
    \label{fig:rm}
\end{figure}

The deconvolved sizes of the 14 identified cores are between 0.04 pc and 0.23 pc, and the derived masses range from 5.4 to 408 M$_{\odot}$ with a mean value of 116.7 M$_{\odot}$. In order to determine whether the selected dust cores have sufficient mass to form massive stars, we try to investigate the relationship between radius and mass. According to \cite{2010ApJ...723L...7K}, if the core mass satisfy this criterion, $M(r)\geq580M_{\odot}(r/pc)^{1.33}$, then they can potentially form massive stars. Figure~\ref{fig:rm} presents a mass versus radius plot of 14 candidate cores, where the solid black line represents the threshold that can be used to separate the regimes under which massive stars can form (above the line) or not (below the line). From Figure~\ref{fig:rm}, we found that 11 of 14 candidate cores lie at or above the threshold, indicating that 78$\%$ of these cores have sufficient density and mass to form massive stars. 
 
The core formation efficiency (CFE) is the ratio of the total core mass to cloud mass, $CFE = M_{\rm cores}/M_{\rm cloud}$, which exhibits the efficiency in gas-to-core convertion. The total mass of the cores in the ridge is 1.6$\times 10^3$ M$_{\odot}$, and the mass of the ridge is 6.7$\times 10^3$ M$_{\odot}$, then we obtained the observed CFE of 24\%. \cite{2007MNRAS.379..663M} found that the CFE in the W3 molecular cloud is 5\%-13\% in the diffuse cloud and 26\%-37\% in the compressed region which agrees well with our result. There are also a number of papers that have estimated core or clump formation efficiency in different environments, and comparing our results with them can gain more useful information about the physical properties of the parent molecular clouds. Recent studies have suggested that the average CFE on Galactic scales is about 8\% \citep{2012MNRAS.422.3178E,2014ApJ...780..173B,2021MNRAS.500..191E}, consistent with the low efficiency found in simulation \citep{2002ApJ...576..870P}. According to the calculation of \cite{2011A&A...529A..41N}, the clumps formation efficiency of the whole W43 molecular complex is about 11\%. However, \cite{2013MNRAS.431.1587E} found that most of the molecular clouds associated with \HII regions have CFE as high as 40\%. The clump formation efficiency obtained in the M16 \HII region is 22$\pm$3\% \citep{2019A&A...627A..27X}. \cite{2020A&A...639A..93F} estimated that the CFE of \HII region RCW 120 varies between 12 and 26\%. Moreover, \cite{2017A&A...605A..35P} concluded that the expansion of early-stage \HII regions create higher CFE. Hence, the best interpretation for the high core formation efficiency is that the feedback from the \HII regions, where the expanding \HII regions compress the molecular clouds to create the high-density structure, or inject energy into the molecular clouds.

\begin{table*}
\tabcolsep 1mm\caption{Physical parameters of two sub-filaments and different predicted values of two fragmentation models.}
\def\temptablewidth{1\textwidth}
\begin{center}
\begin{tabular}{lcccccccccccccc}
\hline\hline
Region & $N_{\rm {H_2}}$ & $T_{\rm {dust}}$ & $\lambda_{\rm observed}$ & $\lambda_{\rm Jeans1}$ & $\lambda_{\rm Jeans2}$ &$\lambda_{\rm cl1}$& $\lambda_{\rm cl2}$&  \\
  &($\times$10$^{4}$ cm$^{-3}$) & (k) & (pc) & (pc) & (pc) & (pc) & (pc) & \\   
    \hline\noalign{\smallskip} 
Sub-filament A & 8 & 15 & 0.51 & 0.11 & 0.37 & 0.31 & 1.38 \\
Sub-filament B & 10 & 15 & 0.45 & 0.09 & 0.33 & 0.28 & 1.24 \\
\noalign{\smallskip}\hline
\end{tabular}\end{center}
{Note: $N_{\rm {H_2}}$ and $T_{\rm {dust}}$ are the average column density and average dust temperature of two sub-filaments. $\lambda_{\rm observed}$ is the average observed spacing of cores. $\lambda_{\rm Jeans1}$ and $\lambda_{\rm Jeans2}$ are the Jeans length under the Jeans fragmentation model dominate by thermal pressure and turbulence, respectively. Similarly, $\lambda_{\rm cl1}$ and $\lambda_{\rm cl2}$ are the fragmentation spacing due to the ``sausage'' instability where the cylinder is supported by the thermal pressure or the turbulent pressure. }
\label{Table:spacing}
\end{table*}

\subsection{Fragmentation of two sub-filaments in AFGL 333-Ridge}
Previous work has shown that filaments are prone to hierarchical fragmentation, creating the structure of different spatial scales \citep{2013ApJ...763...57T,2013A&A...557A.120K,2014MNRAS.439.3275W,2016A&A...587A..47T}. The hierarchical structures in AFGL-333 also indicated fragmentation at different spatial scales such as the ridge, sub-filaments, and dense cores (see in Figure~\ref{fig:position}). In this section, we primarily discuss the fragmentation of sub-filaments in the ridge. The seven cores in the two sub-filaments of AFGL 333-Ridge seem to constitute two necklace-like chains. We estimated that the average density of sub-filament A is $8 \times 10^4$ cm$^{-3}$ and average density of sub-filament B is $1 \times 10^5$ cm$^{-3}$. The average temperature of both sub-filaments is 15 K. While the average observed spacing of dense cores in the two sub-filaments is calculated to be 0.51 pc and 0.45 pc, respectively, based on the central position of cores given by {\it {\it getsources}} outcomes.

The Jeans fragmentation model is considered effective to predict the observed spacing. In an infinite and homogenous medium where Jeans instability dominates, the Jeans length is described as \citep{1902RSPTA.199....1J}, $\lambda = c_s^2(\frac{\pi}{G \rho})^{1/2}$, where G is the gravitational constant, $\rho$ is the mean density ($\rho = \mu m_{H_2}n$, where $n$ is the particle number density). We obtain the Jeans fragmentation length as 0.11 pc and 0.09pc for subfilament A and B, respectively. If turbulence plays a dominant role in the internal pressure of sub-filaments, the $c_s$ should be replaced by velocity dispersion ($\sigma$). Then the predicted value turns out to be 0.37 pc and 0.33 pc. The predicted values from Jeans fragmentation is smaller than the observed spacing (0.45 pc and 0.51 pc) in the sub-filaments, meaning that the Jeans fragmentation mode is not suitable for the sub-filaments.

Furthermore, assuming the sub-filament is an infinitely long static and cylindrical isothermal cloud, the sub-filament will fragment into cores or clumps due to the ``sausage'' instability \citep{1993PASJ...45..551N,2010ApJ...719L.185J,2016MNRAS.456.2041C}. Moreover, in this condition, the clumps or cores are distributed at the characteristic spacing, given by $\lambda_{cl}=22 \upsilon (4 \pi G \rho_c)^{-1/2} $, where $\rho_c$ is the average gas density of the sub-filaments. If the cylinder is supported by the thermal pressure, the $c_s$ should be instead of $\nu$, with a theoretical spacing of 0.31 pc for sub-filament A and 0.28 pc for sub-filament B. Similarly, if the cylinder is supported by the turbulent pressure, the $\nu$ should be replaced by $\sigma$, with the characteristic spacing of 1.38 pc and 1.24 pc for subfilament A and B, respectively. We list all the theoretical values under different fragmentation models in Table~\ref{Table:spacing}. In addition, if the influence of the magnetic field is taken into account, the theoretical core spacing will reduce accordingly \citep{2000MNRAS.311...85F,2016MNRAS.456.2041C}. We find that the observed spacing is larger than the predicted value from cylindrical fragmentation governed by thermal pressure, but smaller than that governed by turbulence. Therefore, we argue that the observed spacing is not caused by a single factor, but by a combination of thermal pressure and turbulence in the cylindrical fragmentation model. Here, our observations may indicate the presence of a multilayered fragmentation process in the molecular cloud. In AFGL 333-Ridge, the collision produces elongated sub-filaments structures, and these supercritical sub-filaments fragment into chains of closely-spaced cores, while some dense cores even have the possibility of further fragmentation.

\section{Conclusion}
\label{sec:conclusions}

The principal conclusions of this study can be summarized as follows:

1. The statistical analysis on velocity dispersion shows that turbulence is the dominant factor in AFGL 333-Ridge. The linear mass of the filament is much greater than the critical values and therefore some other motions must be present to prevent the filament from radially collapse, such as magnetic field or feedback.

2. 14 dense cores were identified using the {\it {\it getsources}} algorithm, including 2 protostellar cores and 12 prestellar cores. The correlation between radius and mass suggests that 78\% of the cores on the ridge are dense and massive enough to form massive stars. The ionization feedback from the \HII regions create a dense structure in the AFGL 333-Ridge, thus may yield a higher core formation efficiency (24$\%$).

3. AFGL 333-Ridge presents different structures of hierarchical fragmentation. The observed spacing between cores embedded in the sub-filament A and sub-filament B is approximately 0.51 pc and 0.45 pc, respectively, which is generally in agreement with the theoretical spacing of cylindrical fragmentation. Moreover, we suggest that the combined effect of thermal pressure and turbulence causes the sub-filaments to fragment into the cores. 
\acknowledgments

We thank the referee for insightful comments that improved the clarity of this manuscript.  This work was  supported by the Youth Innovation Promotion Association of CAS, the National Natural Science Foundation of China (Grant No.11933011), and supported by the Open Project Program of the Key Laboratory of FAST, NAOC, Chinese Academy of Sciences. 

\bibliographystyle{aasjournal}
\bibliography{ref}

\begin{thebibliography}{}
\expandafter\ifx\csname natexlab\endcsname\relax\def\natexlab#1{#1}\fi
\providecommand{\url}[1]{\href{#1}{#1}}
\providecommand{\dodoi}[1]{doi:~\href{http://doi.org/#1}{\nolinkurl{#1}}}
\providecommand{\doeprint}[1]{\href{http://ascl.net/#1}{\nolinkurl{http://ascl.net/#1}}}
\providecommand{\doarXiv}[1]{\href{https://arxiv.org/abs/#1}{\nolinkurl{https://arxiv.org/abs/#1}}}

\bibitem[{{Aguirre} {et~al.}(2011){Aguirre}, {Ginsburg}, {Dunham}, {Drosback},
  {Bally}, {Battersby}, {Bradley}, {Cyganowski}, {Dowell}, {Evans}, {Glenn},
  {Harvey}, {Rosolowsky}, {Stringfellow}, {Walawender}, \&
  {Williams}}]{2011ApJS..192....4A}
{Aguirre}, J.~E., {Ginsburg}, A.~G., {Dunham}, M.~K., {et~al.} 2011, \apjs,
  192, 4, \dodoi{10.1088/0067-0049/192/1/4}

\bibitem[{{Andr{\'e}} {et~al.}(2010){Andr{\'e}}, {Men'shchikov}, {Bontemps},
  {K{\"o}nyves}, {Motte}, {Schneider}, {Didelon}, {Minier}, {Saraceno},
  {Ward-Thompson}, {di Francesco}, {White}, {Molinari}, {Testi}, {Abergel},
  {Griffin}, {Henning}, {Royer}, {Mer{\'\i}n}, {Vavrek}, {Attard},
  {Arzoumanian}, {Wilson}, {Ade}, {Aussel}, {Baluteau}, {Benedettini},
  {Bernard}, {Blommaert}, {Cambr{\'e}sy}, {Cox}, {di Giorgio}, {Hargrave},
  {Hennemann}, {Huang}, {Kirk}, {Krause}, {Launhardt}, {Leeks}, {Le Pennec},
  {Li}, {Martin}, {Maury}, {Olofsson}, {Omont}, {Peretto}, {Pezzuto}, {Prusti},
  {Roussel}, {Russeil}, {Sauvage}, {Sibthorpe}, {Sicilia-Aguilar}, {Spinoglio},
  {Waelkens}, {Woodcraft}, \& {Zavagno}}]{2010A&A...518L.102A}
{Andr{\'e}}, P., {Men'shchikov}, A., {Bontemps}, S., {et~al.} 2010, \aap, 518,
  L102, \dodoi{10.1051/0004-6361/201014666}

\bibitem[{{Battisti} \& {Heyer}(2014)}]{2014ApJ...780..173B}
{Battisti}, A.~J., \& {Heyer}, M.~H. 2014, \apj, 780, 173,
  \dodoi{10.1088/0004-637X/780/2/173}

\bibitem[{{Condon} {et~al.}(1998){Condon}, {Cotton}, {Greisen}, {Yin},
  {Perley}, {Taylor}, \& {Broderick}}]{1998AJ....115.1693C}
{Condon}, J.~J., {Cotton}, W.~D., {Greisen}, E.~W., {et~al.} 1998, \aj, 115,
  1693, \dodoi{10.1086/300337}

\bibitem[{{Contreras} {et~al.}(2016){Contreras}, {Garay}, {Rathborne}, \&
  {Sanhueza}}]{2016MNRAS.456.2041C}
{Contreras}, Y., {Garay}, G., {Rathborne}, J.~M., \& {Sanhueza}, P. 2016,
  \mnras, 456, 2041, \dodoi{10.1093/mnras/stv2796}

\bibitem[{{Dunham} {et~al.}(2008){Dunham}, {Crapsi}, {Evans}, {Bourke},
  {Huard}, {Myers}, \& {Kauffmann}}]{2008ApJS..179..249D}
{Dunham}, M.~M., {Crapsi}, A., {Evans}, Neal~J., I., {et~al.} 2008, \apjs, 179,
  249, \dodoi{10.1086/591085}

\bibitem[{{Eden} {et~al.}(2013){Eden}, {Moore}, {Morgan}, {Thompson}, \&
  {Urquhart}}]{2013MNRAS.431.1587E}
{Eden}, D.~J., {Moore}, T.~J.~T., {Morgan}, L.~K., {Thompson}, M.~A., \&
  {Urquhart}, J.~S. 2013, \mnras, 431, 1587, \dodoi{10.1093/mnras/stt279}

\bibitem[{{Eden} {et~al.}(2012){Eden}, {Moore}, {Plume}, \&
  {Morgan}}]{2012MNRAS.422.3178E}
{Eden}, D.~J., {Moore}, T.~J.~T., {Plume}, R., \& {Morgan}, L.~K. 2012, \mnras,
  422, 3178, \dodoi{10.1111/j.1365-2966.2012.20840.x}

\bibitem[{{Eden} {et~al.}(2021){Eden}, {Moore}, {Plume}, {Rigby}, {Urquhart},
  {Marsh}, {Pe{\~n}aloza}, {Clark}, {Smith}, {Tahani}, {Ragan}, {Thompson},
  {Johnstone}, {Parsons}, \& {Rani}}]{2021MNRAS.500..191E}
{Eden}, D.~J., {Moore}, T.~J.~T., {Plume}, R., {et~al.} 2021, \mnras, 500, 191,
  \dodoi{10.1093/mnras/staa3188}

\bibitem[{{Feh{\'e}r} {et~al.}(2017){Feh{\'e}r}, {Juvela}, {Lunttila},
  {Montillaud}, {Ristorcelli}, {Zahorecz}, \& {T{\'o}th}}]{2017A&A...606A.102F}
{Feh{\'e}r}, O., {Juvela}, M., {Lunttila}, T., {et~al.} 2017, \aap, 606, A102,
  \dodoi{10.1051/0004-6361/201629866}

\bibitem[{{Fiege} \& {Pudritz}(2000)}]{2000MNRAS.311...85F}
{Fiege}, J.~D., \& {Pudritz}, R.~E. 2000, \mnras, 311, 85,
  \dodoi{10.1046/j.1365-8711.2000.03066.x}

\bibitem[{{Figueira} {et~al.}(2020){Figueira}, {Zavagno}, {Bronfman},
  {Russeil}, {Finger}, \& {Schuller}}]{2020A&A...639A..93F}
{Figueira}, M., {Zavagno}, A., {Bronfman}, L., {et~al.} 2020, \aap, 639, A93,
  \dodoi{10.1051/0004-6361/202037713}

\bibitem[{{Griffin} {et~al.}(2010){Griffin}, {Abergel}, {Abreu}, {Ade},
  {Andr{\'e}}, {Augueres}, {Babbedge}, {Bae}, {Baillie}, {Baluteau}, {Barlow},
  {Bendo}, {Benielli}, {Bock}, {Bonhomme}, {Brisbin}, {Brockley-Blatt},
  {Caldwell}, {Cara}, {Castro-Rodriguez}, {Cerulli}, {Chanial}, {Chen},
  {Clark}, {Clements}, {Clerc}, {Coker}, {Communal}, {Conversi}, {Cox},
  {Crumb}, {Cunningham}, {Daly}, {Davis}, {de Antoni}, {Delderfield}, {Devin},
  {di Giorgio}, {Didschuns}, {Dohlen}, {Donati}, {Dowell}, {Dowell}, {Duband},
  {Dumaye}, {Emery}, {Ferlet}, {Ferrand}, {Fontignie}, {Fox}, {Franceschini},
  {Frerking}, {Fulton}, {Garcia}, {Gastaud}, {Gear}, {Glenn}, {Goizel},
  {Griffin}, {Grundy}, {Guest}, {Guillemet}, {Hargrave}, {Harwit}, {Hastings},
  {Hatziminaoglou}, {Herman}, {Hinde}, {Hristov}, {Huang}, {Imhof}, {Isaak},
  {Israelsson}, {Ivison}, {Jennings}, {Kiernan}, {King}, {Lange}, {Latter},
  {Laurent}, {Laurent}, {Leeks}, {Lellouch}, {Levenson}, {Li}, {Li},
  {Lilienthal}, {Lim}, {Liu}, {Lu}, {Madden}, {Mainetti}, {Marliani}, {McKay},
  {Mercier}, {Molinari}, {Morris}, {Moseley}, {Mulder}, {Mur}, {Naylor},
  {Nguyen}, {O'Halloran}, {Oliver}, {Olofsson}, {Olofsson}, {Orfei}, {Page},
  {Pain}, {Panuzzo}, {Papageorgiou}, {Parks}, {Parr-Burman}, {Pearce},
  {Pearson}, {P{\'e}rez-Fournon}, {Pinsard}, {Pisano}, {Podosek}, {Pohlen},
  {Polehampton}, {Pouliquen}, {Rigopoulou}, {Rizzo}, {Roseboom}, {Roussel},
  {Rowan-Robinson}, {Rownd}, {Saraceno}, {Sauvage}, {Savage}, {Savini},
  {Sawyer}, {Scharmberg}, {Schmitt}, {Schneider}, {Schulz}, {Schwartz},
  {Shafer}, {Shupe}, {Sibthorpe}, {Sidher}, {Smith}, {Smith}, {Smith},
  {Spencer}, {Stobie}, {Sudiwala}, {Sukhatme}, {Surace}, {Stevens}, {Swinyard},
  {Trichas}, {Tourette}, {Triou}, {Tseng}, {Tucker}, {Turner}, {Vaccari},
  {Valtchanov}, {Vigroux}, {Virique}, {Voellmer}, {Walker}, {Ward}, {Waskett},
  {Weilert}, {Wesson}, {White}, {Whitehouse}, {Wilson}, {Winter}, {Woodcraft},
  {Wright}, {Xu}, {Zavagno}, {Zemcov}, {Zhang}, \&
  {Zonca}}]{2010A&A...518L...3G}
{Griffin}, M.~J., {Abergel}, A., {Abreu}, A., {et~al.} 2010, \aap, 518, L3,
  \dodoi{10.1051/0004-6361/201014519}

\bibitem[{{Inutsuka} \& {Miyama}(1992)}]{1992ApJ...388..392I}
{Inutsuka}, S.-I., \& {Miyama}, S.~M. 1992, \apj, 388, 392,
  \dodoi{10.1086/171162}

\bibitem[{{Inutsuka} \& {Miyama}(1997)}]{1997ApJ...480..681I}
{Inutsuka}, S.-i., \& {Miyama}, S.~M. 1997, \apj, 480, 681,
  \dodoi{10.1086/303982}

\bibitem[{{Jackson} {et~al.}(2010){Jackson}, {Finn}, {Chambers}, {Rathborne},
  \& {Simon}}]{2010ApJ...719L.185J}
{Jackson}, J.~M., {Finn}, S.~C., {Chambers}, E.~T., {Rathborne}, J.~M., \&
  {Simon}, R. 2010, \apjl, 719, L185, \dodoi{10.1088/2041-8205/719/2/L185}

\bibitem[{{Jeans}(1902)}]{1902RSPTA.199....1J}
{Jeans}, J.~H. 1902, Philosophical Transactions of the Royal Society of London
  Series A, 199, 1, \dodoi{10.1098/rsta.1902.0012}

\bibitem[{{Kainulainen} {et~al.}(2013){Kainulainen}, {Ragan}, {Henning}, \&
  {Stutz}}]{2013A&A...557A.120K}
{Kainulainen}, J., {Ragan}, S.~E., {Henning}, T., \& {Stutz}, A. 2013, \aap,
  557, A120, \dodoi{10.1051/0004-6361/201321760}

\bibitem[{{Kauffmann} \& {Pillai}(2010)}]{2010ApJ...723L...7K}
{Kauffmann}, J., \& {Pillai}, T. 2010, \apjl, 723, L7,
  \dodoi{10.1088/2041-8205/723/1/L7}

\bibitem[{{K{\"o}nyves} {et~al.}(2015){K{\"o}nyves}, {Andr{\'e}},
  {Men'shchikov}, {Palmeirim}, {Arzoumanian}, {Schneider}, {Roy}, {Didelon},
  {Maury}, {Shimajiri}, {Di Francesco}, {Bontemps}, {Peretto}, {Benedettini},
  {Bernard}, {Elia}, {Griffin}, {Hill}, {Kirk}, {Ladjelate}, {Marsh}, {Martin},
  {Motte}, {Nguy{\^e}n Luong}, {Pezzuto}, {Roussel}, {Rygl}, {Sadavoy},
  {Schisano}, {Spinoglio}, {Ward-Thompson}, \& {White}}]{2015A&A...584A..91K}
{K{\"o}nyves}, V., {Andr{\'e}}, P., {Men'shchikov}, A., {et~al.} 2015, \aap,
  584, A91, \dodoi{10.1051/0004-6361/201525861}

\bibitem[{{K{\"o}nyves} {et~al.}(2020){K{\"o}nyves}, {Andr{\'e}},
  {Arzoumanian}, {Schneider}, {Men'shchikov}, {Bontemps}, {Ladjelate},
  {Didelon}, {Pezzuto}, {Benedettini}, {Bracco}, {Di Francesco}, {Goodwin},
  {Rygl}, {Shimajiri}, {Spinoglio}, {Ward-Thompson}, \&
  {White}}]{2020A&A...635A..34K}
{K{\"o}nyves}, V., {Andr{\'e}}, P., {Arzoumanian}, D., {et~al.} 2020, \aap,
  635, A34, \dodoi{10.1051/0004-6361/201834753}

\bibitem[{{Krumholz} \& {McKee}(2008)}]{2008Natur.451.1082K}
{Krumholz}, M.~R., \& {McKee}, C.~F. 2008, \nat, 451, 1082,
  \dodoi{10.1038/nature06620}

\bibitem[{{Liang} {et~al.}(2021){Liang}, {Xu}, {Xu}, \&
  {Wang}}]{2021ApJ...913...14L}
{Liang}, X., {Xu}, J.-L., {Xu}, Y., \& {Wang}, J.-J. 2021, \apj, 913, 14,
  \dodoi{10.3847/1538-4357/abf1eb}

\bibitem[{{McKee} \& {Ostriker}(2007)}]{2007ARA&A..45..565M}
{McKee}, C.~F., \& {Ostriker}, E.~C. 2007, \araa, 45, 565,
  \dodoi{10.1146/annurev.astro.45.051806.110602}

\bibitem[{{Men'shchikov} {et~al.}(2012){Men'shchikov}, {Andr{\'e}}, {Didelon},
  {Motte}, {Hennemann}, \& {Schneider}}]{2012A&A...542A..81M}
{Men'shchikov}, A., {Andr{\'e}}, P., {Didelon}, P., {et~al.} 2012, \aap, 542,
  A81, \dodoi{10.1051/0004-6361/201218797}

\bibitem[{{Moore} {et~al.}(2007){Moore}, {Bretherton}, {Fujiyoshi}, {Ridge},
  {Allsopp}, {Hoare}, {Lumsden}, \& {Richer}}]{2007MNRAS.379..663M}
{Moore}, T.~J.~T., {Bretherton}, D.~E., {Fujiyoshi}, T., {et~al.} 2007, \mnras,
  379, 663, \dodoi{10.1111/j.1365-2966.2007.11941.x}

\bibitem[{{Nakamura} {et~al.}(1993){Nakamura}, {Hanawa}, \&
  {Nakano}}]{1993PASJ...45..551N}
{Nakamura}, F., {Hanawa}, T., \& {Nakano}, T. 1993, \pasj, 45, 551

\bibitem[{{Nakano} {et~al.}(2017){Nakano}, {Soejima}, {Chibueze}, {Nagayama},
  {Omodaka}, {Handa}, {Sunada}, {Kamezaki}, \& {Burns}}]{2017PASJ...69...16N}
{Nakano}, M., {Soejima}, T., {Chibueze}, J.~O., {et~al.} 2017, \pasj, 69, 16,
  \dodoi{10.1093/pasj/psw120}

\bibitem[{{Nguyen Luong} {et~al.}(2011){Nguyen Luong}, {Motte}, {Schuller},
  {Schneider}, {Bontemps}, {Schilke}, {Menten}, {Heitsch}, {Wyrowski},
  {Carlhoff}, {Bronfman}, \& {Henning}}]{2011A&A...529A..41N}
{Nguyen Luong}, Q., {Motte}, F., {Schuller}, F., {et~al.} 2011, \aap, 529, A41,
  \dodoi{10.1051/0004-6361/201016271}

\bibitem[{{Padoan} \& {Nordlund}(2002)}]{2002ApJ...576..870P}
{Padoan}, P., \& {Nordlund}, {\r{A}}. 2002, \apj, 576, 870,
  \dodoi{10.1086/341790}

\bibitem[{{Palmeirim} {et~al.}(2013){Palmeirim}, {Andr{\'e}}, {Kirk},
  {Ward-Thompson}, {Arzoumanian}, {K{\"o}nyves}, {Didelon}, {Schneider},
  {Benedettini}, {Bontemps}, {Di Francesco}, {Elia}, {Griffin}, {Hennemann},
  {Hill}, {Martin}, {Men'shchikov}, {Molinari}, {Motte}, {Nguyen Luong},
  {Nutter}, {Peretto}, {Pezzuto}, {Roy}, {Rygl}, {Spinoglio}, \&
  {White}}]{2013A&A...550A..38P}
{Palmeirim}, P., {Andr{\'e}}, P., {Kirk}, J., {et~al.} 2013, \aap, 550, A38,
  \dodoi{10.1051/0004-6361/201220500}

\bibitem[{{Palmeirim} {et~al.}(2017){Palmeirim}, {Zavagno}, {Elia}, {Moore},
  {Whitworth}, {Tremblin}, {Traficante}, {Merello}, {Russeil}, {Pezzuto},
  {Cambr{\'e}sy}, {Baldeschi}, {Bandieramonte}, {Becciani}, {Benedettini},
  {Buemi}, {Bufano}, {Bulpitt}, {Butora}, {Carey}, {Costa}, {Deharveng}, {Di
  Giorgio}, {Eden}, {Hajnal}, {Hoare}, {Kacsuk}, {Leto}, {Marsh}, {M{\`e}ge},
  {Molinari}, {Molinaro}, {Noriega-Crespo}, {Schisano}, {Sciacca}, {Trigilio},
  {Umana}, \& {Vitello}}]{2017A&A...605A..35P}
{Palmeirim}, P., {Zavagno}, A., {Elia}, D., {et~al.} 2017, \aap, 605, A35,
  \dodoi{10.1051/0004-6361/201629963}

\bibitem[{{Poglitsch} {et~al.}(2010){Poglitsch}, {Waelkens}, {Geis},
  {Feuchtgruber}, {Vandenbussche}, {Rodriguez}, {Krause}, {Renotte}, {van
  Hoof}, {Saraceno}, {Cepa}, {Kerschbaum}, {Agn{\`e}se}, {Ali}, {Altieri},
  {Andreani}, {Augueres}, {Balog}, {Barl}, {Bauer}, {Belbachir}, {Benedettini},
  {Billot}, {Boulade}, {Bischof}, {Blommaert}, {Callut}, {Cara}, {Cerulli},
  {Cesarsky}, {Contursi}, {Creten}, {De Meester}, {Doublier}, {Doumayrou},
  {Duband}, {Exter}, {Genzel}, {Gillis}, {Gr{\"o}zinger}, {Henning},
  {Herreros}, {Huygen}, {Inguscio}, {Jakob}, {Jamar}, {Jean}, {de Jong},
  {Katterloher}, {Kiss}, {Klaas}, {Lemke}, {Lutz}, {Madden}, {Marquet},
  {Martignac}, {Mazy}, {Merken}, {Montfort}, {Morbidelli}, {M{\"u}ller},
  {Nielbock}, {Okumura}, {Orfei}, {Ottensamer}, {Pezzuto}, {Popesso},
  {Putzeys}, {Regibo}, {Reveret}, {Royer}, {Sauvage}, {Schreiber}, {Stegmaier},
  {Schmitt}, {Schubert}, {Sturm}, {Thiel}, {Tofani}, {Vavrek}, {Wetzstein},
  {Wieprecht}, \& {Wiezorrek}}]{2010A&A...518L...2P}
{Poglitsch}, A., {Waelkens}, C., {Geis}, N., {et~al.} 2010, \aap, 518, L2,
  \dodoi{10.1051/0004-6361/201014535}

\bibitem[{{Takahashi} {et~al.}(2013){Takahashi}, {Ho}, {Teixeira}, {Zapata}, \&
  {Su}}]{2013ApJ...763...57T}
{Takahashi}, S., {Ho}, P. T.~P., {Teixeira}, P.~S., {Zapata}, L.~A., \& {Su},
  Y.-N. 2013, \apj, 763, 57, \dodoi{10.1088/0004-637X/763/1/57}

\bibitem[{{Teixeira} {et~al.}(2016){Teixeira}, {Takahashi}, {Zapata}, \&
  {Ho}}]{2016A&A...587A..47T}
{Teixeira}, P.~S., {Takahashi}, S., {Zapata}, L.~A., \& {Ho}, P.~T.~P. 2016,
  \aap, 587, A47, \dodoi{10.1051/0004-6361/201526807}

\bibitem[{{Wang} {et~al.}(2014){Wang}, {Zhang}, {Testi}, {van der Tak}, {Wu},
  {Zhang}, {Pillai}, {Wyrowski}, {Carey}, {Ragan}, \&
  {Henning}}]{2014MNRAS.439.3275W}
{Wang}, K., {Zhang}, Q., {Testi}, L., {et~al.} 2014, \mnras, 439, 3275,
  \dodoi{10.1093/mnras/stu127}

\bibitem[{{Ward-Thompson} {et~al.}(2007){Ward-Thompson}, {Andr{\'e}},
  {Crutcher}, {Johnstone}, {Onishi}, \& {Wilson}}]{2007prpl.conf...33W}
{Ward-Thompson}, D., {Andr{\'e}}, P., {Crutcher}, R., {et~al.} 2007, in
  Protostars and Planets V, ed. B.~{Reipurth}, D.~{Jewitt}, \& K.~{Keil}, 33

\bibitem[{{Xu} {et~al.}(2019){Xu}, {Zavagno}, {Yu}, {Liu}, {Xu}, {Yuan},
  {Zhang}, {Zhang}, {Zhang}, {Ning}, \& {Ju}}]{2019A&A...627A..27X}
{Xu}, J.-L., {Zavagno}, A., {Yu}, N., {et~al.} 2019, \aap, 627, A27,
  \dodoi{10.1051/0004-6361/201935024}

\end{thebibliography}


\end{document}